\documentclass[aps,prc,reprint,smsmath,amssymb,superscriptaddress]{revtex4-1}

\usepackage{graphicx}
\usepackage{dcolumn}
\usepackage{bm}
\usepackage{amsmath}
\usepackage{lineno}

\begin{document}
\title{First measurement of coherent $\phi$-meson photoproduction from helium-4\\near threshold}

\author{T.~Hiraiwa}
\affiliation{Research Center for Nuclear Physics, Osaka University, Ibaraki, Osaka 567-0047, Japan}
\affiliation{Department of Physics, Kyoto University, Kyoto 606-8502, Japan}
\author{M.~Yosoi}
\affiliation{Research Center for Nuclear Physics, Osaka University, Ibaraki, Osaka 567-0047, Japan}
\author{M.~Niiyama}
\affiliation{Department of Physics, Kyoto University, Kyoto 606-8502, Japan}
\author{Y.~Morino}
\affiliation{High Energy Accelerator Organization (KEK), Tsukuba, Ibaraki 305-0801, Japan}
\author{Y.~Nakatsugawa} 
\affiliation{Institute of High Energy Physics, Chinese Academy of Sciences, Beijing 100049, China}
\author{M.~Sumihama}
\affiliation{Department of Education, Gifu University, Gifu 501-1193, Japan}
\author{D.~S.~Ahn}
\affiliation{RIKEN, Nishina Center for Accelerator-Based Science, Wako, Saitama 351-0198, Japan}
\author{J.~K.~Ahn}
\affiliation{Department of Physics, Korea University, Seoul 02841, Republic of Korea}
\author{W.~C.~Chang}
\affiliation{Institute of Physics, Academia Sinica, Taipei 11529, Taiwan}
\author{J.~Y.~Chen}
\affiliation{Light Source Division, National Synchrotron Radiation Research Center, Hsinchu 30076, Taiwan}
\author{S.~Dat\'{e}}
\affiliation{Japan Synchrotron Radiation Research Institute, Sayo, Hyogo 679-5143, Japan}
\author{H.~Fujimura}
\affiliation{Wakayama Medical University, Wakayama 641-8509, Japan}
\author{S.~Fukui}
\affiliation{Research Center for Nuclear Physics, Osaka University, Ibaraki, Osaka 567-0047, Japan}
\author{K.~Hicks}
\affiliation{Department of Physics and Astronomy, Ohio University, Athens, Ohio 45701, USA}
\author{T.~Hotta}
\affiliation{Research Center for Nuclear Physics, Osaka University, Ibaraki, Osaka 567-0047, Japan}
\author{S.~H.~Hwang}
\affiliation{Korea Research Institute of Standards and Science, Daejeon 34113, Republic of Korea}
\author{T.~Ishikawa}
\affiliation{Research Center for Electron Photon Science, Tohoku University, Sendai, Miyagi 982-0826, Japan}
\author{Y.~Kato}
\affiliation{Department of Physics and Astrophysics, Nagoya University, Nagoya, Aichi 464-8602, Japan}
\author{H.~Kawai}
\affiliation{Department of Physics, Chiba University, Chiba 263-8522, Japan} 
\author{H.~Kohri}
\author{Y.~Kon}
\affiliation{Research Center for Nuclear Physics, Osaka University, Ibaraki, Osaka 567-0047, Japan}
\author{P.~J.~Lin}
\affiliation{Institute of Physics, Academia Sinica, Taipei 11529, Taiwan}
\author{Y.~Maeda}
\affiliation{Proton Therapy Center, Fukui Prefectural Hospital, Fukui 910-8526, Japan}
\author{M. Miyabe}
\affiliation{Research Center for Electron Photon Science, Tohoku University, Sendai, Miyagi 982-0826, Japan}
\author{K.~Mizutani}
\affiliation{Department of Physics, Kyoto University, Kyoto 606-8502, Japan}
\author{N.~Muramatsu}
\affiliation{Research Center for Electron Photon Science, Tohoku University, Sendai, Miyagi 982-0826, Japan}
\author{T.~Nakano}
\author{Y.~Nozawa}
\affiliation{Research Center for Nuclear Physics, Osaka University, Ibaraki, Osaka 567-0047, Japan}
\author{Y.~Ohashi}
\affiliation{Japan Synchrotron Radiation Research Institute, Sayo, Hyogo 679-5143, Japan}
\author{T.~Ohta}
\affiliation{Department of Radiology, The University of Tokyo Hospital, Tokyo 113-8655, Japan}
\author{M.~Oka}
\affiliation{Research Center for Nuclear Physics, Osaka University, Ibaraki, Osaka 567-0047, Japan}
\author{C.~Rangacharyulu}
\affiliation{Department of Physics and Engineering Physics, University of Saskatchewan, Saskatoon, Saskatchewan S7N 5E2, Canada}
\author{S.~Y.~Ryu}
\affiliation{Research Center for Nuclear Physics, Osaka University, Ibaraki, Osaka 567-0047, Japan}
\author{T.~Saito}
\affiliation{Department of Physics, Chiba University, Chiba 263-8522, Japan} 
\author{T.~Sawada}
\affiliation{Physics Department, University of Michigan, Michigan 48109-1040, USA}
\author{H.~Shimizu}
\affiliation{Research Center for Electron Photon Science, Tohoku University, Sendai, Miyagi 982-0826, Japan}
\author{E.~A.~Strokovsky}
\affiliation{Joint Institute for Nuclear Research, Dubna, Moscow Region, 142281, Russia}
\affiliation{Research Center for Nuclear Physics, Osaka University, Ibaraki, Osaka 567-0047, Japan}
\author{Y.~Sugaya}
\affiliation{Research Center for Nuclear Physics, Osaka University, Ibaraki, Osaka 567-0047, Japan}
\author{K.~Suzuki}
\affiliation{Department of Physics, Osaka University, Toyonaka, Osaka 560-0043, Japan}
\author{A.~O.~Tokiyasu}
\affiliation{Research Center for Electron Photon Science, Tohoku University, Sendai, Miyagi 982-0826, Japan}
\author{T.~Tomioka}
\affiliation{Department of Physics, Chiba University, Chiba 263-8522, Japan} 
\author{T.~Tsunemi}
\affiliation{Department of Physics, Kyoto University, Kyoto 606-8502, Japan}
\author{M.~Uchida}
\affiliation{Department of Physics, Tokyo Institute of Technology, Tokyo 152-8551, Japan}
\author{T.~Yorita}
\affiliation{Research Center for Nuclear Physics, Osaka University, Ibaraki, Osaka 567-0047, Japan}

\collaboration{LEPS Collaboration}
\noaffiliation

\date{\today}

\begin{abstract}
The differential cross sections and  decay angular distributions for coherent $\phi$-meson photoproduction from helium-4 have been measured for the first time at forward angles with linearly polarized photons
in the energy range $E_{\gamma} = \text{1.685--2.385 GeV}$. Thanks to the target with spin-parity $J^{P} = 0^{+}$, unnatural-parity exchanges are absent, and thus natural-parity exchanges can be
investigated clearly. The decay asymmetry with respect to photon polarization is shown to be very close to the maximal value. This ensures the strong dominance~($> 94\%$) of natural-parity exchanges in this 
reaction. To evaluate the contribution from natural-parity exchanges to the forward cross section~($\theta = 0^\circ$) for the $\gamma p \rightarrow \phi p$ reaction near the threshold, the energy dependence of
the forward cross section~($\theta = 0^\circ$) for the $\gamma {^{4}\text{He}} \rightarrow \phi {^{4}\text{He}}$ reaction was analyzed. The comparison to $\gamma p \rightarrow \phi p$ data suggests that enhancement
of the forward cross section arising from natural-parity exchanges, and/or destructive interference between natural-parity and unnatural-parity exchanges is needed in the $\gamma p \rightarrow \phi p$ reaction near
the threshold.
\end{abstract}

\maketitle

\section{Introduction}
The $\phi$-meson photoproduction offers rich information on gluonic interactions at low energies. Because of almost pure $s\bar{s}$ components of the $\phi$-meson, meson exchanges
in its interactions with nucleons are suppressed by the Okubo-Zweig-Iizuka rule, and multi-gluon exchanges are expected to be dominant. The slow rise of the total
cross section with the energy $\sqrt{s}$ can be well understood by the $t$-channel exchange of gluonic objects with the vacuum quantum numbers, known as the Pomeron trajectory in the Regge
phenomenology~\cite{regge}, in the framework of vector meson dominance~\cite{photopro}. The Pomeron trajectory has been discussed in connection with a glueball trajectory with
$J^{PC} = 2^{++},~4^{++},~\cdots$, etc.~\cite{tensGB1, PomGB, tensGB2}, but it is still an open question what the physical particles lying on the Pomeron trajectory are. While the Pomeron exchange
has successfully described the common features of diffractive hadron-hadron and photon-hadron scatterings at high energies~\cite{diffrac1,diffrac2,diffrac3}, its applicability to low energies is not
completely clear~\cite{philowE1,philowE2}. In the other hadronic reactions such as $pp$ collisions or photoproduction with flavor changing such as pion or kaon production, it is difficult to study the
Pomeron exchange at low energies because meson exchanges become significant. Therefore, the $\phi$-meson photoproduction is unique in studying the Pomeron exchange at low energies~\cite{expPom}
and searching for a new glueball-associated trajectory, i.e. a daughter Pomeron trajectory~\cite{dautPom}, as inspired by the scalar glueball~($J^{PC}=0^{++}$, $M^{2} \sim 3 ~ \text{GeV}^{2}$) predicted by
lattice QCD calculations~\cite{latticeQCD1,latticeQCD2}. 

The LEPS Collaboration measured the $\gamma p \rightarrow \phi p$ reaction near the threshold at forward angles~\cite{LEPSphip,LEPSint}, where the $t$-channel Pomeron exchange is expected to be dominant.
The energy dependence of the forward cross section~($\theta = 0^\circ$) shows a local maximum around $E_{\gamma} \sim 2 ~ \text{GeV}$, which contradicts a monotonic behavior as a Pomeron exchange
model predicts. Such a behavior was also observed by CLAS~\cite{CLASphip,CLASphip_neut}, whereas the data were obtained by extrapolating from the large scattering angle region. Recent measurements by LEPS
extended the maximal beam energy from $2.4 ~ \text{GeV}$ to $2.9 ~ \text{GeV}$ and have confirmed an excess from the monotonic curve of a model prediction~\cite{LEPS3gev}.  Several theoretical models have been
proposed so far~\cite{philowE2,theor_bump_couple,*[Erratum:]theor_bump_couple_erra,theor_bump_sreso,*[Erratum:]theor_bump_sreso_erra,theor_bump_couple_2,theor_bump_diquark}, but no conclusive interpretation has been
obtained yet. From measurements of the $\phi \rightarrow K^{+}K^{-}$ decay angular distributions with linearly polarized photons~\cite{LEPSphip,WC_SDM}, unnatural-parity exchanges such as the $\pi$ and $\eta$
exchanges are known to have a certain contribution~($\sim 30 \%$) near the threshold. 

A coherent photoproduction with an isoscalar target is very useful for studying the Pomeron exchange at low energies since the isovector $\pi$ exchange, which is a dominant meson exchange process, is
forbidden~\cite{coh_phi_omega_fromD,phi_fromD}. The LEPS data for the coherent $\gamma d \rightarrow \phi d$ reaction~\cite{LEPScohphi_fromD} shows that a Pomeron exchange model including small contribution of
the $\eta$ exchange~\cite{phi_fromD} underestimates the energy dependence of the forward cross section~($\theta = 0^\circ$). 

In this article, we present the first measurement of the differential cross sections and decay angular distributions for the coherent $\gamma {^{4}\text{He}} \rightarrow \phi {^{4}\text{He}}$ reaction
at forward angles near the threshold with linearly polarized photons. This reaction has advantages compared to the $\gamma d$ reaction: First, thanks to the $0^{+}$ target, this reaction completely eliminates unnatural-parity
exchanges since a $0^{+}$ particle cannot emit an unnatural-parity particle, remaining unchanged in spin and parity, due to spin-parity conservation. Second, owing to the large separation energy of helium-4 nuclei, the
coherent production events could be cleanly separated from the incoherent ones, even better than what is with deuterium target. Accordingly, we can investigate natural-parity exchanges such as the Pomeron and
multi-gluon exchanges at low energies with better accuracies.

\section{Experiment and analysis}
The experiment was carried out at the SPring-8 facility using the LEPS spectrometer~\cite{sumi_Kprod}. Linearly polarized photons were produced via the backward Compton scattering between UV-laser photons with
a wavelength of $355~ \text{nm}$ and 8-GeV electrons in the storage ring~\cite{LEPbeam}. The photon energy was determined by the momentum analysis of  the recoil electrons with tagging counters.
The photon energy resolution~($\sigma$) was 13.5 MeV for all energies. The degree of photon polarization varied with photon energy; 69\% at $E_{\gamma} = 1.685 ~ \text{GeV}$, and 92\% at
$E_{\gamma} = 2.385 ~ \text{GeV}$.  The systematic uncertainty in the polarization degree was estimated to be less than 0.1\%.  The tagged photons irradiated a liquid helium-4 target with
a length of 15 cm. The integrated flux of the tagged photons was $4.6 \times 10^{12}$. The systematic uncertainty of the photon flux was estimated to be 3\%. Produced charged particles were detected at forward angles,
and their momenta were analyzed by the LEPS spectrometer. The momentum resolution~($\sigma$) of the spectrometer was 0.9\% in ${\delta p}/p$ for typical $\text{1-GeV}/c$ particles. More details about the
experimental setup can be found in Ref.~\cite{tpc_setup}.

\begin{figure}[htb]
\includegraphics[clip,width=8.5cm]{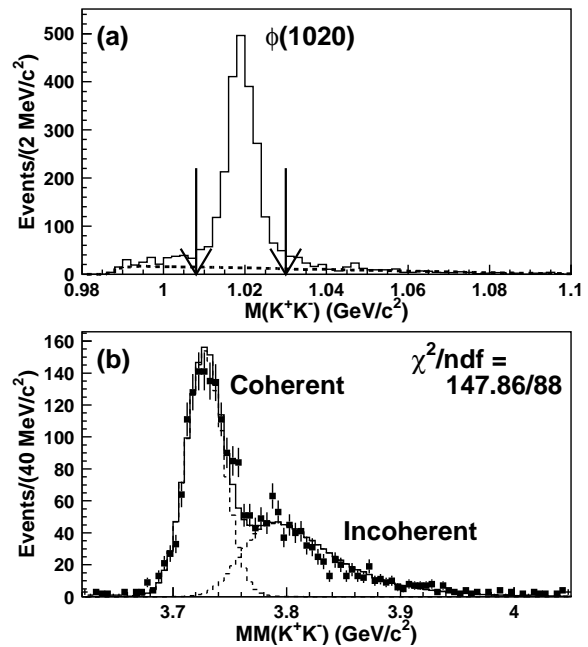}%
 \caption{\label{fig:m_mmkk} (a) Invariant mass spectrum for $K^{+}K^{-}$ pairs. The dashed curve shows the MC-simulated background. The arrows show cut points for selecting the $\phi$-meson events.
 	(b) Missing mass spectrum for the ${^{4}\text{He}}(\gamma, K^{+}K^{-})X$ reaction after selecting the $\phi$-meson events. The solid histogram shows the fit result with two MC templates for the
 		coherent and incoherent processes~(dashed histograms). }
 \end{figure}

The production of  $\phi$-mesons was identified by detecting $K^{+}K^{-}$ tracks from the $\phi \rightarrow K^{+}K^{-}$ decay. $K^{+}K^{-}$ tracks were selected according to the reconstructed mass-squared
and charge by the spectrometer with a $4\sigma$ cut, where $\sigma$ is the momentum-dependent resolution of the reconstructed mass-squared.  The contamination of pions due to particle misidentifications was
reduced to a negligible level by requiring the missing mass of the ${^{4}\text{He}}(\gamma, K^{+}K^{-})X$ reaction to be above $3.62 ~ \text{GeV}/c^{2}$. The $K^{+}K^{-}$ pairs produced inside the target were selected by
imposing a cut on the $z$-positions of the reconstructed vertices of $K^{+}K^{-}$ pairs. Under this cut, the contamination from materials other than the target was estimated to be 2\% with empty-target data.
Figure~\ref{fig:m_mmkk}(a) shows the invariant mass spectrum for the $K^{+}K^{-}$ pairs~[$\text{M}(K^{+}K^{-})$]. A clear signal for $\phi$-mesons was observed on small background contribution from the non-resonant
$K^{+}K^{-}$ production. Note that the quasi-free $K^{+}\Lambda(1520)$ production followed by the $\Lambda(1520) \rightarrow K^{-}p$ decay was found to be negligible at small momentum transfers $|t|$ of our
interest~(${-t} < 0.2 ~ \text{GeV}^{2}$). The $\phi$-meson yields including both coherent and incoherent processes were estimated by fitting invariant mass spectra with Monte Carlo~(MC) templates. The spectral shapes
for the $\phi$-meson and non-resonant $K^{+}K^{-}$ events were reproduced by GEANT3~\cite{g3}-based MC simulations, where the geometrical acceptance, the photon energy resolution, the momentum resolution, and
the detector efficiencies were implemented. The background level under the $\phi$-meson signal was estimated to be 1-15\%, depending on the photon energy and the momentum transfer. 

The coherent events were disentangled from the incoherent events by fitting missing mass spectra for the ${^{4}\text{He}}(\gamma, K^{+}K^{-})X$ reaction~[$\text{MM}(K^{+}K^{-})$] after selecting the
$\phi$-meson events as $1.008 < \text{M}(K^{+}K^{-}) < 1.030 ~ \text{GeV}/c^{2}$~[Fig.~\ref{fig:m_mmkk}(b)].  A clear peak for the coherent $\gamma {^{4}\text{He}} \rightarrow \phi  {^{4}\text{He}}$
reaction was observed around $\text{MM}(K^{+}K^{-}) \approx 3.73 ~ \text{GeV}/c^{2}$, corresponding to the mass of helium-4 nuclei. The spectral shapes for the coherent and incoherent processes were
reproduced by the MC simulations. The missing mass $\text{MM}(K^{+}K^{-})$ resolution~($\sigma$) was estimated to be 14-17~$\text{MeV}/c^{2}$, which was consistent with estimates from hydrogen-target data.

To reproduce the line shape of the $\text{MM}(K^{+}K^{-})$ spectra for the incoherent process, the Fermi motion and off-shell effects of the target nucleon inside a helium-4 nucleus were simulated as follows:
For the off-shell correction, we adopted the first approach in Ref.~\cite{LEPScohphi_fromD}. The Fermi momenta of the target nucleon were taken from the numerical results of variational Monte Carlo
calculations for the helium-4 wave function~\cite{momDist4He}. Moreover, following Ref.~\cite{LEPScohphi_fromD}, the energy dependence of the forward cross section~($\theta = 0^\circ$) for the $\phi$-meson
photoproduction from off-shell nucleons as well as the differential cross section $d\sigma/dt$ was also taken into account.  

Systematic uncertainties due to contamination from events other than the coherent ones were estimated by considering additional processes, in the $\text{MM}(K^{+}K^{-})$ fits, such as
\begin{equation}
\begin{split}
 \gamma  + {\text{\textquoteleft}t\text{\textquoteright}} &\rightarrow \phi + t, \\
 \gamma + {\text{\textquoteleft}d\text{\textquoteright}} &\rightarrow \phi + d,  
\end{split}
\label{eq:sem_coh}
\end{equation}
\noindent
where ${\text{\textquoteleft}t\text{\textquoteright}}~({\text{\textquoteleft}d\text{\textquoteright}})$ stands for the triton~(deuteron) wave function in helium-4 nuclei. The off-shell effects of the triton and deuteron clusters
inside a helium-4 nucleus were simulated in the same manner as that for the incoherent process. Their Fermi momenta were taken from Ref.~\cite{momDist4He}.

The acceptance of the LEPS spectrometer including all the detector efficiencies and the analysis efficiency was calculated by using the MC simulation. The detector efficiencies were evaluated from the data channel by channel,
and were taken into account position-dependently in the MC simulation. The simulation was iterated so as to reproduce the measured differential cross section $d\sigma/dt$ and decay angular distributions.  The validity of the
acceptance calculation as well as the normalization of the photon flux was checked with hydrogen-target data taken in the same period, by comparing the differential cross sections of other reactions with the previous LEPS
measurements~\cite{LEPSphip,sumi_Kprod,sumi_pi0prod}.

\section{Decay Angular Distribution}
First, we present the $\phi \rightarrow K^{+}K^{-}$ decay angular distributions in the Gottfried-Jackson frame. The three-dimensional decay angular distribution, $W(\cos\Theta, \Phi, \Psi)$, with linearly polarized photons, as a
function of the polar~($\Theta$) and azimuthal~($\Phi$) angles of the $K^{+}$ and the azimuthal angle~($\Psi$) of the photon polarization with respect to the production plane, are parametrized by the nine spin density matrix
elements~($\rho^{i}_{jk}$) and the degree of photon polarization~($P_{\gamma}$)~\cite{classicalSDM}.  Following Ref.~\cite{titovSDM}, one obtains five one-dimensional decay angular distributions:
\begin{equation}
\begin{split}
W(\cos\Theta) &= \frac{3}{2} \left[ \frac{1}{2}(1-\rho^{0}_{00})\sin^{2}\Theta + \rho^{0}_{00}\cos^{2}\Theta \right], \\
W(\Phi)           &= \frac{1}{2\pi}(1 - 2\text{Re}\rho^{0}_{1-1}\cos2\Phi), \\
W(\Phi-\Psi)   &= \frac{1}{2\pi} \left[ 1 + 2P_{\gamma}{\overline{\rho}^{1}_{1-1}}\cos2(\Phi-\Psi) \right], \\
W(\Phi+\Psi)   &= \frac{1}{2\pi} \left[ 1 + 2P_{\gamma}\Delta_{1-1}\cos2(\Phi+\Psi) \right], \\
W(\Psi)           &=  1 - P_{\gamma}(2\rho^{1}_{11} + \rho^{1}_{00})\cos2\Psi,
\end{split}
\label{eq:decayangle}
\end{equation}
\noindent
where ${\overline{\rho}^{1}_{1-1}} \equiv (\rho^{1}_{1-1} - \text{Im}\rho^{2}_{1-1})/2$ and $\Delta_{1-1} \equiv (\rho^{1}_{1-1} + \text{Im}\rho^{2}_{1-1})/2$. These distributions were measured at
$0 < |t| - |t|_{\text{min}} < 0.2 ~ \text{GeV}^{2}$ for two photon energy regions~(E1: $1.985 < E_{\gamma} < 2.185 ~ \text{GeV}$, E2: $2.185 < E_{\gamma} < 2.385 ~ \text{GeV}$), where sufficient statistics were obtained.
Here, $|t|_{\text{min}}$ is the minimum $|t|$ for a helium-4 nucleus.

\begin{table*}[htb]
\caption{\label{tab:SDMels} Extracted spin density matrix elements for the E1 and E2 regions. The first uncertainties are statistical and the second systematic.}
\begin{ruledtabular}
\begin{tabular}{crrrrr}
$E_{\gamma}$ range (GeV) & \multicolumn{1}{c}{$\rho^{0}_{00}$} & \multicolumn{1}{c}{$\text{Re}\rho^{0}_{1-1}$} & \multicolumn{1}{c}{${\overline{\rho}^{1}_{1-1}}$} & \multicolumn{1}{c}{$\Delta_{1-1}$} &
	\multicolumn{1}{c}{$2\rho^{1}_{11} + \rho^{1}_{00}$} \\ \hline
  (E1)~1.985 -- 2.185 & $-0.015 \pm 0.016 ^{+0.000}_{-0.002}$ & $0.116 \pm 0.030 ^{+0.000}_{-0.006}$ & $0.454 \pm 0.024 ^{+0.014}_{-0.000}$ & $-0.111 \pm 0.033 ^{+0.006}_{-0.000}$ & $0.132 \pm 0.066 ^{+0.000}_{-0.033}$ \\
  (E2)~2.185 -- 2.385 & $0.015 \pm 0.012 ^{+0.002}_{-0.000}$ & $0.054 \pm 0.020 ^{+0.000}_{-0.004}$ & $0.436 \pm 0.014 ^{+0.004}_{-0.000}$ & $-0.034 \pm 0.017 ^{+0.009}_{-0.000}$ & $0.074 \pm 0.041^{+0.011}_{-0.000}$ \\
\end{tabular}
\end{ruledtabular}
\end{table*}

\begin{figure}[htb]
\includegraphics[clip,width=8.5cm]{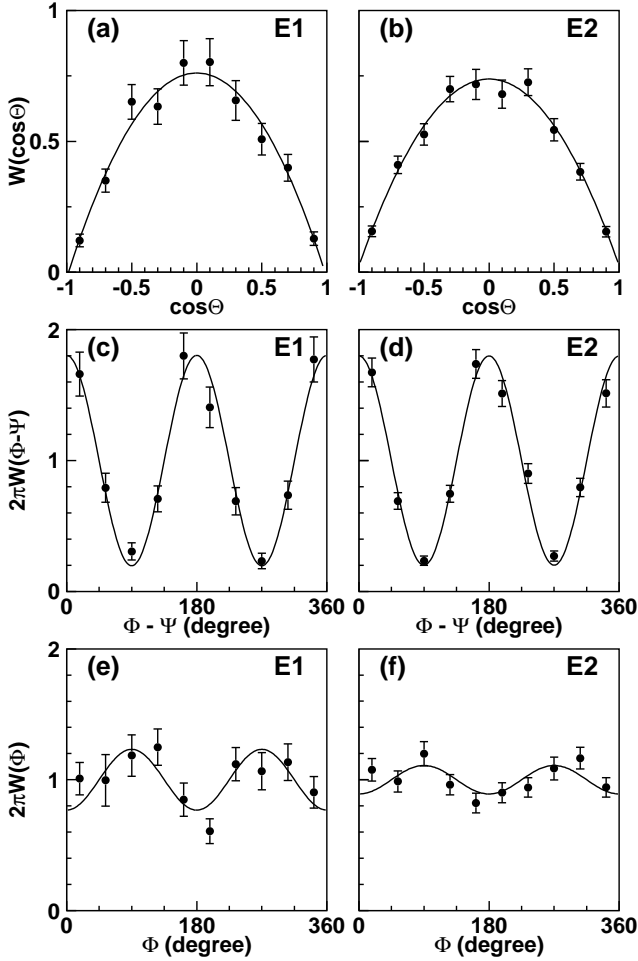}%
 \caption{\label{fig:decayangle} Acceptance-corrected decay angular distribution for the $\gamma {^{4}\text{He}}$ reaction. (a) $W(\cos\Theta)$ for E1 and (b) E2. (c) $W(\Phi-\Psi)$ for E1 and (d) E2. (e) $W(\Phi)$ for E1 and
 	(f) E2. The error bars represent statistical ones only. The solid curves are the fits to the data by Eqs.~(\ref{eq:decayangle}).}
 \end{figure}

Figures~\ref{fig:decayangle}(a) and (b) show the distribution $W(\cos\Theta)$. The extracted spin density matrix elements are summarized in Table~\ref{tab:SDMels}. For both the E1 and E2 regions, $\rho^{0}_{00}$ is consistent 
with zero, which is the same as those for the $\gamma p$ and $\gamma d$ reactions~\cite{LEPSphip,WC_SDM}.  This indicates the dominance of helicity-conserving processes in $t$-channel.

The decay asymmetry, ${\overline{\rho}^{1}_{1-1}}$, is obtained from $W(\Phi-\Psi)$~[Figs.~\ref{fig:decayangle}(c) and (d)]. It reflects the relative contribution of natural-parity and unnatural-parity exchanges, and gives
${+0.5}~({-0.5})$ for pure natural-parity~(unnatural-parity) exchanges when helicity-conservation holds~\cite{classicalSDM,titovSDM}. As shown in Figs.~\ref{fig:decayangle}(c) and (d), quite large oscillations were observed in
$W(\Phi-\Psi)$, and therefore a finite bin size could affect the extracted values of ${\overline{\rho}^{1}_{1-1}}$ by directly using Eq.~(\ref{eq:decayangle}). To avoid such finite bin size effects, a fit chi-square, $\chi^{2}$, was
defined as
\begin{equation}
\begin{split}
\chi^{2}({\overline{\rho}^{1}_{1-1}},\alpha) &= \sum_{i=1}^{N}\frac{(\hat{O}_{i} - \alpha \hat{E}_{i})^{2}}{\sigma_{i}^{2}}, \\
\hat{E}_{i} &= \frac{1}{{\Delta x}} \int_{\bar{x}_{i} - \frac{1}{2}{\Delta x}}^{\bar{x}_{i} + \frac{1}{2}{\Delta x}} W(\Phi-\Psi;=x) \: dx,
\end{split}
\label{eq:chi2}
\end{equation} 
\noindent
where $N$ denotes the number of data points~(bins), $\hat{O}_{i}$ is the number of counts in the $i$-th bin, $\alpha$ denotes an overall normalization factor being a free parameter, $\sigma_{i}$ is the statistical error in the $i$-th bin,
${\Delta x}$ is the bin size, and $\bar{x}_{i}$ is the mean value of the $i$-th bin. We found ${\overline{\rho}^{1}_{1-1}}$ to be very close to ${+0.5}$ for both the E1 and E2 regions, indicating almost pure natural-parity exchanges. 
However, ${\overline{\rho}^{1}_{1-1}}$ sizably deviates from ${+0.5}$. This can be understood by the contribution from double helicity-flip transitions from the incident photon to the outgoing $\phi$-meson~\cite{titovSDM}. In fact,
a rather large oscillation of $W(\Phi)$ was observed in the E1 region~[Fig.~\ref{fig:decayangle}(e)], giving the spin density matrix element of $\text{Re}\rho^{0}_{1-1} \sim 0.11$. This means that the interference of helicity-nonflip
and double helicity-flip amplitudes has a non-zero value~\cite{SLAC_Vmeson_linearpol}.  A non-zero $\text{Re}\rho^{0}_{1-1}$ was also observed in the $\gamma p$~\cite{LEPSphip,LEPS3gev,WC_SDM} and $\gamma d$
reactions~\cite{WC_SDM}. In particular, the $\text{Re}\rho^{0}_{1-1}$ obtained here exhibits a similar energy dependence to that in Ref.~\cite{LEPSphip}. Note that the deviation of ${\overline{\rho}^{1}_{1-1}}$ is not due to the 
contamination from the incoherent events with ${\overline{\rho}^{1}_{1-1}} \approx 0.25$~\cite{LEPSincphi_fromD} because such a deviation does not disappear when a tight mass cut, $\text{MM}(K^{+}K^{-}) < 3.72 ~ \text{GeV}/c^{2}$,
is applied.

\section{Differential Cross Section}
\begin{figure}[htb]
\includegraphics[clip,width=8.5cm]{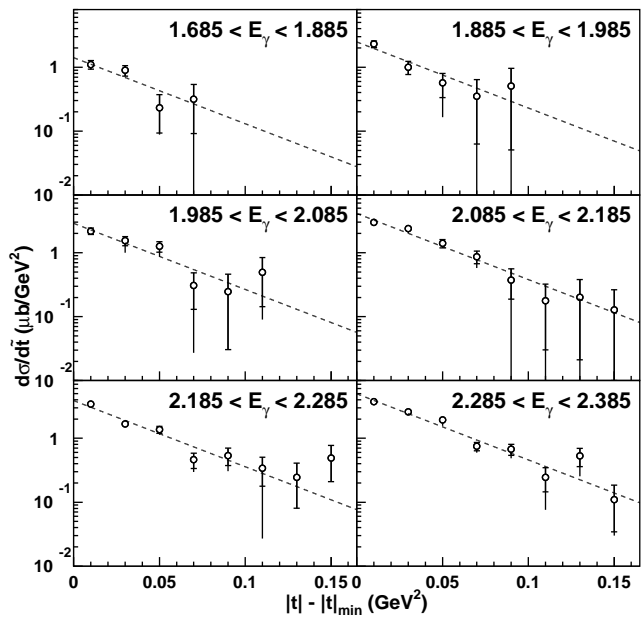}%
 \caption{\label{fig:dsdtall} Differential cross section $d\sigma/d\tilde{t}$ for the $\gamma {^{4}\text{He}}$ reaction. The smaller error bars on the vertical axis represent a statistical error, whereas the larger ones represent a
 	sum of 	the statistical and systematic errors in quadrature. The dashed curves show the fit results by an exponential function with the common slope $b = 23.81~ \text{GeV}^{-2}$.}
 \end{figure}

The differential cross sections as a function of momentum transfer $\tilde{t} ~ (\equiv |t| - |t|_{\text{min}})$, $d\sigma/d\tilde{t}$, were measured in the energy range $E_{\gamma} = \text{1.685--2.385 GeV}$~(Fig.~\ref{fig:dsdtall}).
A strong forward-peaking behavior of $d\sigma/d\tilde{t}$ predominantly comes from the helium-4 form factor. To extract the slope of $d\sigma/d\tilde{t}$, the fit was performed with an exponential function;
$(d\sigma/dt)_{0}^{\gamma {^{4}\text{He}}}\exp(-b\tilde{t})$, where $(d\sigma/dt)_{0}^{\gamma {^{4}\text{He}}}$ is $d\sigma/d\tilde{t}$ at $t = -|t|_{\text{min}}$ and $b$ the slope parameter. No strong energy dependence of the slope
$b$ was found, and the common slope $b$ was determined to be $23.81 \pm 0.95(\text{stat}) ~ ^{+5.16}_{-0.00}(\text{sys}) ~ \text{GeV}^{-2}$. The slope $b$ is consistent with a simple estimate from a single-scattering
assumption~\cite{phi_fromD}, in which the slope $b$ is approximately expressed as $b \approx b_{0} + b_{F}$, where $b_{0}$ is the slope for the elementary $\gamma p$ reaction~($3.38 \pm 0.23 ~ \text{GeV}^{-2}$~\cite{LEPSphip}) 
and $b_{F}$ the slope for the squared charge form factor of helium-4 nuclei~($\approx 22 ~ \text{GeV}^{-2}$~\cite{FF_alpha}). The slope $b$ is also quite reasonable compared with that for other elastic scattering of a hadron off
helium-4 in the diffractive regime~\cite{ela_alpha_p,ela_pi_alpha}. Note that the systematic error on the slope $b$ comes solely from the assumption of the additional processes~[Eq.~(\ref{eq:sem_coh})] in the $\text{MM}(K^{+}K^{-})$
fits.

\begin{figure}[htb]
\includegraphics[clip,width=8.5cm]{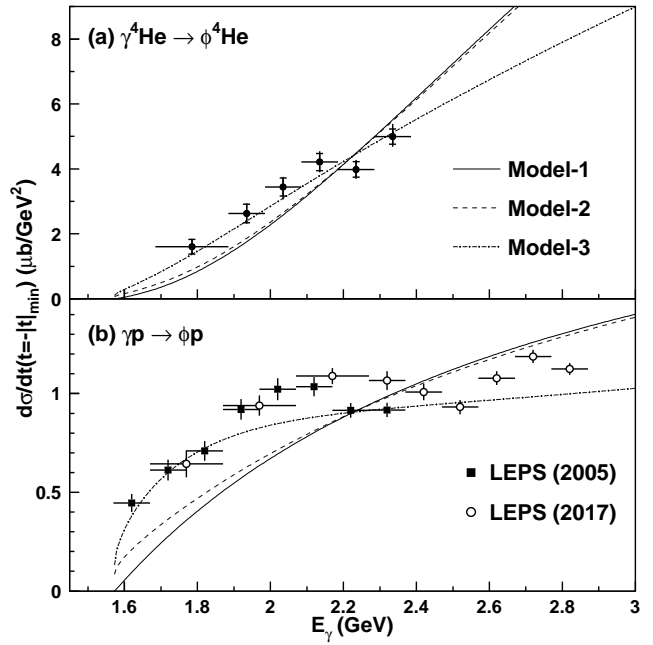}%
 \caption{\label{fig:comp_dsdt0} (a) Energy dependence of $(d\sigma/dt)_{0}^{\gamma {^{4}\text{He}}}$ with the common slope $b = 23.81 ~ \text{GeV}^{-2}$. The meanings of the error bars are the same as those in Fig.~\ref{fig:dsdtall}.
	The solid, dashed and dash-dotted curves are the best fits for model-1, -2 and -3~(explained in the text), respectively. (b) Contribution from natural-parity exchanges to the forward cross section~($\theta = 0^\circ $) for the
	$\gamma p$ reaction with model-1~(solid), -2~(dashed) and -3~(dash-dotted). The experimental data for the $\gamma p$ reaction are represented by filled squares~\cite{LEPSphip} and open circles~\cite{LEPS3gev}.}
\end{figure}

Figure~\ref{fig:comp_dsdt0}(a) shows the energy dependence of $(d\sigma/dt)_{0}^{\gamma {^{4}\text{He}}}$ with the common slope $b = 23.81 ~ \text{GeV}^{-2}$. The differences between the intercepts
$(d\sigma/dt)_{0}^{\gamma {^{4}\text{He}}}$ with the fixed~(common) and variable~(energy-dependent) slopes were found to be within the statistical errors. Also, the systematic errors on $(d\sigma/dt)_{0}^{\gamma {^{4}\text{He}}}$
due to the assumption of the additional processes~[Eq.~(\ref{eq:sem_coh})] in the $\text{MM}(K^{+}K^{-})$ fits were found to be small~(1.5--6.5\%) compared with the statistical ones, though these are reflected in the final results.

As we shall see, it is difficult to discuss the precise energy dependence of the forward cross section~($\theta = 0^\circ$) for the $\gamma p$ reaction arising from natural-parity
exchanges~[$\equiv (d\sigma/dt)_{0}^{\gamma p; \text{NP}}$, where ``NP'' denotes the contribution from natural-parity exchanges.] directly from the $\gamma {^{4}\text{He}}$ data due to the helium-4 form factor. To evaluate
the contribution from natural-parity exchanges to the $\gamma p$ reaction, we constructed  three different models for the energy dependence of $(d\sigma/dt)_{0}^{\gamma p; \text{NP}}$, where their overall strengths are
unknown and to be determined. The first one~(model-1) is simple; that is, $(d\sigma/dt)_{0}^{\gamma p; \text{NP}}$ grows with the energy as $(k_{\phi}/k_{\gamma})^{2}$~\cite{est_dsdt0_exp}, where $k_{\phi}$~($k_{\gamma}$) is the
3-momentum of $\phi$-mesons~(photons) in the center-of-mass frame. The second one~(model-2) is a conventional Pomeron exchange model as in Ref.~\cite{phi_fromD}. The third one~(model-3) describes a threshold
enhancement in the energy dependence of $(d\sigma/dt)_{0}^{\gamma p; \text{NP}}$. This could be realized by modifying the conventional Pomeron exchange model, and/or a manifestation of additional natural-parity exchanges
near the threshold. For model-3, we used the Pomeron and daughter Pomeron exchange model in Ref.~\cite{philowE2}. The relative contribution from the daughter Pomeron exchange was adjusted so as to fit available
low-energy $\gamma p$ data~\cite{LEPSphip,CLASphip_neut,LEPS3gev}.

A theoretical calculation for the coherent $\gamma d$ reaction has been done by A. I. Titov \textit{et al.}~\cite{phi_fromD}, in which they describe the forward cross section by using the amplitudes for the elementary $\gamma p$
reaction and the deuteron form factor. Similarly, $(d\sigma/dt)_{0}^{\gamma {^{4}\text{He}}}$ is described by using the charge form factor for helium-4~($|F_{C}|^{2}$)~\cite{FF_alpha} as
\begin{equation}
\left(\frac{d\sigma}{dt}\right)_{0}^{\gamma {^{4}\text{He}}} = 16|F_{C}|^{2} \left(\frac{d\sigma}{dt}\right)_{0}^{\gamma p; \text{NP}}.
\label{eq:cs0}
\end{equation}
\noindent
Here, $|F_{\text{C}}|^{2}$ is evaluated at $t= -|t|_{\text{min}}$. To fix the overall strengths for the above models, we used this relation in the 
fit to the $\gamma {^{4}\text{He}}$ data with the overall strengths as free parameters. The best fits for model-1, -2 and -3 are depicted in Fig.~\ref{fig:comp_dsdt0}(a) as solid, dashed and dash-dotted curves, respectively. The
$\chi^{2}/\text{ndf}$'s are $48.8/5$, $39.8/5$ and $10.2/5$ for model-1, -2 and -3, respectively.

Figure~\ref{fig:comp_dsdt0}(b) shows the contribution from natural-parity exchanges to the forward cross section~($\theta = 0^\circ$) for the $\gamma p$ reaction with each model, together with the experimental data
by LEPS~\cite{LEPSphip,LEPS3gev}. Model-1 and -2 gave similar results, and we found both the curves to be slightly above the data points for $E_{\gamma} > 2.4 ~ \text{GeV}$. On the other hand, the experimental data on the
decay asymmetry ${\overline{\rho}^{1}_{1-1}}$~\cite{LEPS3gev} shows a sizable 20-30\% contribution from unnatural-parity exchanges to the forward cross section for $2.4 < E_{\gamma} < 2.9 ~ \text{GeV}$. This suggests that 
destructive interference between natural-parity and unnatural-parity exchanges is needed to explain both the measurements of the forward cross section and decay asymmetry. In contrast to model-1 and -2, model-3 describes
the experimental data fairly well. For $E_{\gamma} > 1.9 ~ \text{GeV}$, we found the curve to be below the data by $\sim 20\%$, except for a few data points. This can be compensated by the observed 20-40\% contribution from
unnatural-parity exchanges~\cite{LEPSphip,WC_SDM,LEPS3gev}. In such case, large interference effects between natural-parity and unnatural-parity exchanges are not needed, which is compatible with our current understanding
that the interference effect between the Pomeron and $\pi$ exchanges would be small~\cite{photopro,philowE2}. Note that destructive interference between natural-parity and unnatural-parity exchanges is also needed for
$E_{\gamma} < 1.9 ~ \text{GeV}$ because simply adding the unnatural-parity contribution~($\sim 30\%$) overestimates the experimental data.

\section{Conclusion}
In conclusion, we presented the first measurement of the differential cross sections and decay angular distributions for coherent $\phi$-meson photoproduction from helium-4 at forward angles with linearly polarized
photons in the energy range $E_{\gamma} = \text{1.685-2.385 GeV}$. With the elimination of unnatural-parity exchanges, this reaction provides a unique and clean way of investigating natural-parity exchanges
in $\phi$-meson photoproduction at low energies. The measurement of ${\overline{\rho}^{1}_{1-1}}$ demonstrates the strong dominance~($> 94\%$) of natural-parity exchanges in this reaction. Three different models were
constructed for describing the contribution from natural-parity exchanges to the forward cross section~($\theta=0^\circ$) for the $\gamma p$ reaction near the threshold, and their overall strengths were determined from
the present data. The comparison of them to available $\gamma p$ data suggests that enhancement of the forward cross section arising from natural-parity exchanges, and/or destructive interference between natural-parity
and unnatural-parity exchanges is needed in the $\gamma p$ reaction near the threshold. Further theoretical and experimental efforts are of great help for revealing the underlying reaction mechanisms in the $\phi$-meson
photoproduction at low energies.

\begin{acknowledgments}
The authors thank the staff at SPring-8 for supporting the LEPS experiment. We thank A. I. Titov, A. Hosaka and H. Nagahiro for fruitful discussions. The experiment was performed at the BL33LEP of SPring-8 with the approval of
the Japan Synchrotron Radiation Research Institute~(JASRI) as a contract beamline~(Proposal No. BL33LEP/6001). This work was supported in part by the Ministry of Education, Science, Sports and Culture of Japan, the National
Science Council of the Republic of China~(Taiwan), the National Science Foundation~(USA) and the National Research Foundation~(Korea).  
\end{acknowledgments}

\bibliography{prc_coh_phi_4he}

\end{document}